\documentclass[preprint,epsfig,showpacs,preprintnumbers,amsmath,amssymb]{revtex4-2}
\usepackage[dvips]{graphicx} \usepackage{amsmath}
\usepackage{color}
\definecolor{textcolor}{cmyk}{0,0,0,1}
\definecolor{magenta}{rgb}{1,0,1}
\definecolor{green}{rgb}{0,1,0}
\definecolor{red}{rgb}{1,0,0}
\begin{document}
%\draft
\title{
Restored quantum size effects of Pb overlayers at high coverages.}
\author{A. Ayuela$^*$, E. Ogando$^{\dag}$, and N. Zabala$^{*\dag}$}\affiliation{$^*$  Donostia International Physics
 Center (DIPC) and Unidad F\'isica de Materiales, Centro Mixto CSIC-UPV/EHU, 20018
Donostia, Spain\\ $^{\dag}$ Elektrizitatea eta
Elektronika Saila, Zientzia eta Teknologia Fakultatea UPV-EHU 644 P.K., 48080 Bilbao,
Spain}
% \date{\today}

\begin{abstract}
Abnormally large  stability of Pb nanostructures grown  on metallic or
semiconductor substrates  has been observed even for  heights of about
30 monolayers.  Using both  density functional theory calculations and
analytical models,  we demonstrate that  the stability at  even higher
coverages ($N>30$  MLs) is supported  by an extra second  quantum beat
pattern  in the  energetics of  the metal  film as  a function  of the
number   of  atomic  layers.   This  pattern   is  triggered   by  the
butterfly-like  shape  of the  Fermi  surface  of  lead in  the  (111)
direction and supports the detection of stable magic islands of higher
heights than measured up to now.
\end{abstract}

\maketitle

The literature of the last  twenty years presents many examples of new
structures of nanometer scale,  showing a different behaviour from the
bulk due to quantum size  effects (QSE's). The quantum oscillations of
the physical properties of  nanostructures decay in amplitude as their
size increases,  but an  open general question  is the  convergence of
this behavior  towards the bulk.   Modulations in the  oscillations of
quantum origin
%of several systems 
are observed  for the stability of atomic  clusters \cite{genzken} and
nanowires  \cite{yanson1999,yanson2000,yanson2001}.  These modulations
are also present in the growth of thin films, such as Pb layers, which
is our topic.

As  a  result  of  the  electronic confinement  perpendicular  to  the
substrate, QSE's show  up during the growth of  Pb metallic thin films
\cite{luh,upton,czoschke,floreano,aballe,saito}, when the thickness is
comparable with the Fermi wavelength $\lambda_F$.  Pb nanoislands over
Cu(111)  \cite{otero}  or Si(111)  \cite{hupalo}  have revealed  their
preference for bi-layer growth in the [1 1 1] orientation.  The origin
of the  "magic" height selection  can be understood qualitatively
with  a simple  picture  of  electrons confined  in  a potential  well
\cite{zhang,schulte}.  When  the energy of  the films is  displayed with
the number of MLs, Fig. 1 (a), a modulated pattern or quantum
beat structure is obtained, as the interlayer spacing is approximately
$3/4\lambda      _F$       \cite{edu1,edu2}.       Previous      works
\cite{luh,upton,czoschke,floreano,aballe,saito,otero,hupalo,zhang,schulte,edu1,edu2}
have used  only a single value  of the Fermi wave  vector of Pb.
Nevertheless, a quantitative agreement with the measured magic heights
requires more sophisticated models.

Extra  modulations or  quantum beats  given by  extra nestings  of the
Fermi surface have been shown  in other context, such as sandwiched Co
magnetic     layers      with     a     nonmagnetic      Cu     spacer
\cite{parkin,johnson,bruno1991,edwards}.     Both    theoretical   and
experimental works, discussed at length the role of the second Cu wave
vector  nesting  on  the  magnetic  coupling  between  the  sandwiched
magnetic layers.  However, the role of a second wave vector nesting on
the stability during the growth of layers has not been addressed yet.

In this  paper we explore  larger sizes (up  to 60 MLs) of  Pb thin
films than usually  researched in the literature by  using the Density
Functional Theory  (DFT) and including the atomic  structure. To study
the stability of the films we have calculated the energy as a function
of the  number of Pb MLs.   Our main ab-initio result  is displayed in
Fig. \ref{fig:beating_ab_initio}.   We find that  the oscillatory part
of the energy as a function of the number of ML's has a second quantum
beat  structure that emerges  from the  nesting of  two Pb  Fermi wave
vectors,  as  we are  explaining  later.   This  second pattern  affects
the amplitude  of these  oscillations and  explains the
stability up to the calculated thickness of 60 MLs. We are able to fit
recent  measurements  up  to   35  MLs  \cite{czoschke}  with  a  more
sophisticated model  that assumes two $k_F$  values. Nevertheless, the
effects of this second beat  pattern emerge clearly just close to this
thickness of 35 MLs, so  for a more  extensive comparison, experiments  
for even higher coverages would be helpful.

\begin{figure}
\includegraphics[width=0.8\columnwidth]{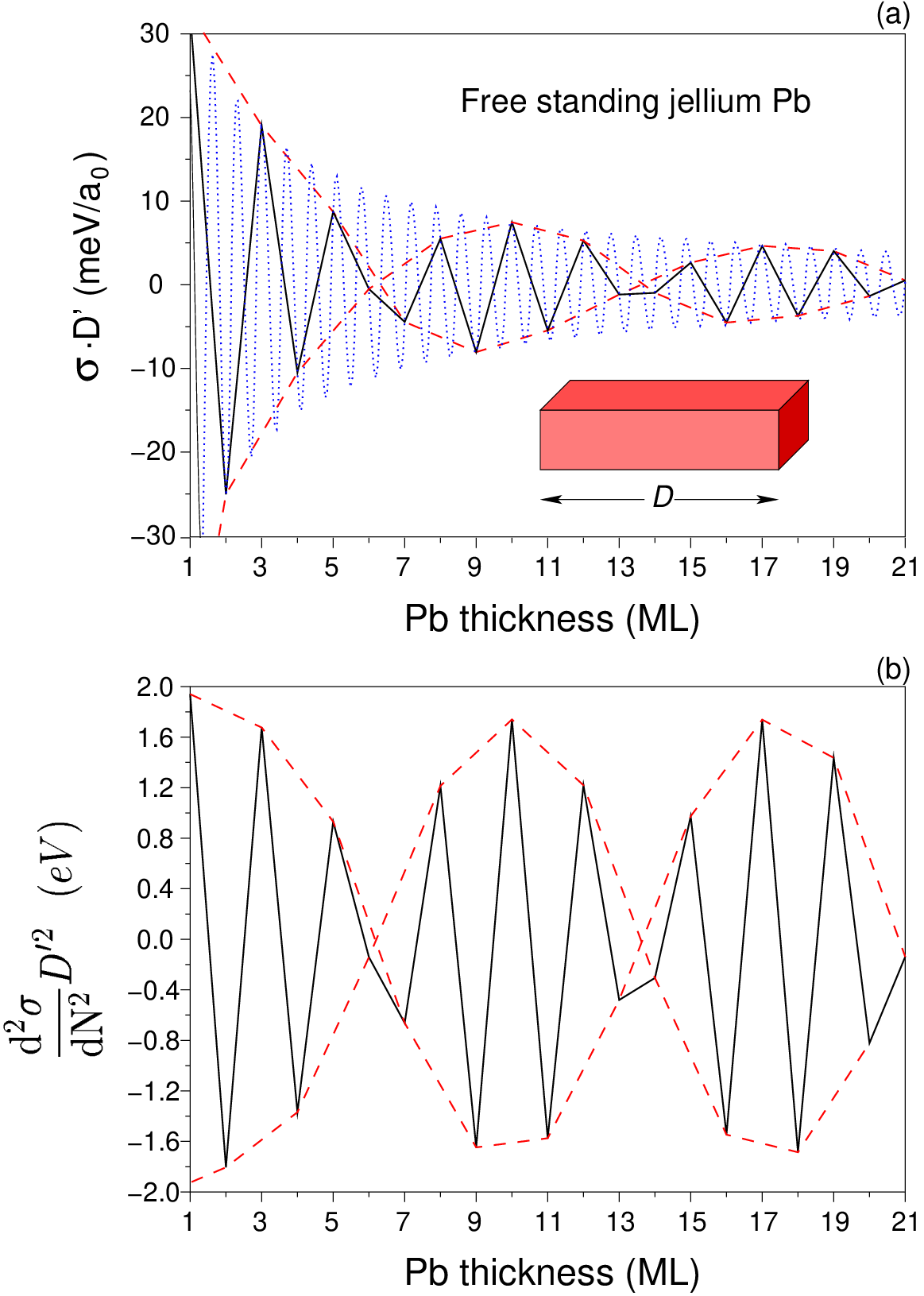}
\caption{{(\it  color on  line)}  (a) Oscillating  part  of the  total
energy per surface area($\sigma$)  multiplied by the Pb slab effective
thickness  $D'$ for a  free standing  Pb, and  calculated by  means of
self-consistent jellium calculations  \cite{edu1}.  The dotted line is
a function of the continuous thickness, and the full line connects the
values of completed  MLs.  The dashed lines guide the  eye to show the
(first) beat pattern.  (b) Second derivative of $\sigma$ multiplied
by the square of the effective thickness (black line) and its envelope
function (blue line).  }
\label{fig:jell-ana}
\end{figure}

Let us first summarise the  main features for Pb islands, successfully
modeled   with   the  uniform   jellium   model   in  previous   works
\cite{edu1,edu2}. The oscillating part of  the energy as a function of
the  Pb thickness is  given in  Fig.  \ref{fig:jell-ana}(a)  with full
line. The minima  of the curve correspond to  the theoretically stable
islands.  Alternatively, one can look at (minus) the second derivative
of the energy versus the thickness, the number of ML's (N), as
shown in Fig.  \ref{fig:jell-ana}(b).  The valleys are associated with
the most stable sizes  \cite{upton,czoschke}, as also studied in wires
and clusters \cite{alonso}.  The second derivative has been multiplied
by the square of the  effective thickness \cite{edu2} ($D'^2$) to show
the damping of the amplitude.

These  oscillations and  their decay  using the  jellium  model assume
implicitly  only one  Fermi wavelength,  i.e. the  Fermi surface  is a
circle in  the directions perpendicular  to the surface film.  In this
work  we   address  the   question  concerning  stability   at  larger
thicknesses ($N>30$ MLs)  of the Pb slabs, when we  allow for a second
critical spanning vector of the  Fermi surface.  In the context of the
present investigation, the second  Fermi wavelength arises because the
Fermi surface in the (111) direction is not perfectly circular but has
a butterfly  shape \cite{otero,anderson}, as sketched in  the inset of
Fig. \ref{fig:beating_ana}.

Jellium calculations  describe the  amplitude of the  oscillations and
their phase  for a  spherical Fermi surface.   Therefore, in  order to
analyse the  effect of  the crystal structure  with a  realistic Fermi
surface we have calculated with DFT ab-initio methods, the energy as a
function  of thickness for  the free  standing Pb  slabs grown  in the
(111) direction.  The  calculations have been done with  the VASP code
\cite{vasp}  by  using the  generalized  gradient approximation  (GGA)
\cite{perdew} for the exchange-correlation potential and the projector
augmented-wave method (PAW)  \cite{kresse}. Details on our calculation
are  given in  Ref.  \cite{details}. We assume bulk  distances
between atoms \cite{wei}.  For our purpose, it is  unnecessary to relax
the atomic  positions because self-compression effects  at the surface
produce negligible changes in the energy \cite{materzanini}. Stress at
the interface or interaction with the substrate produce a shift on the
oscillating structure or slight  differences affecting only to thickness 
up to ten ML. But  they do  not change the  qualitative behaviour  of the
oscillations.  Therefore  we  neglect   the  substrate  in  this  study
\cite{edu1}. As we are interested in  the stability of  the slabs, we
have calculated first the total energy  as a function of the number of
atomic layers $N$ up to $N=60$,  and then we have evaluated the second
derivatives of the energy versus  the thickness as done before in Fig.
1 (b).  Our result is displayed in Fig. \ref{fig:beating_ab_initio}.

\begin{figure}
\includegraphics[width=\columnwidth]{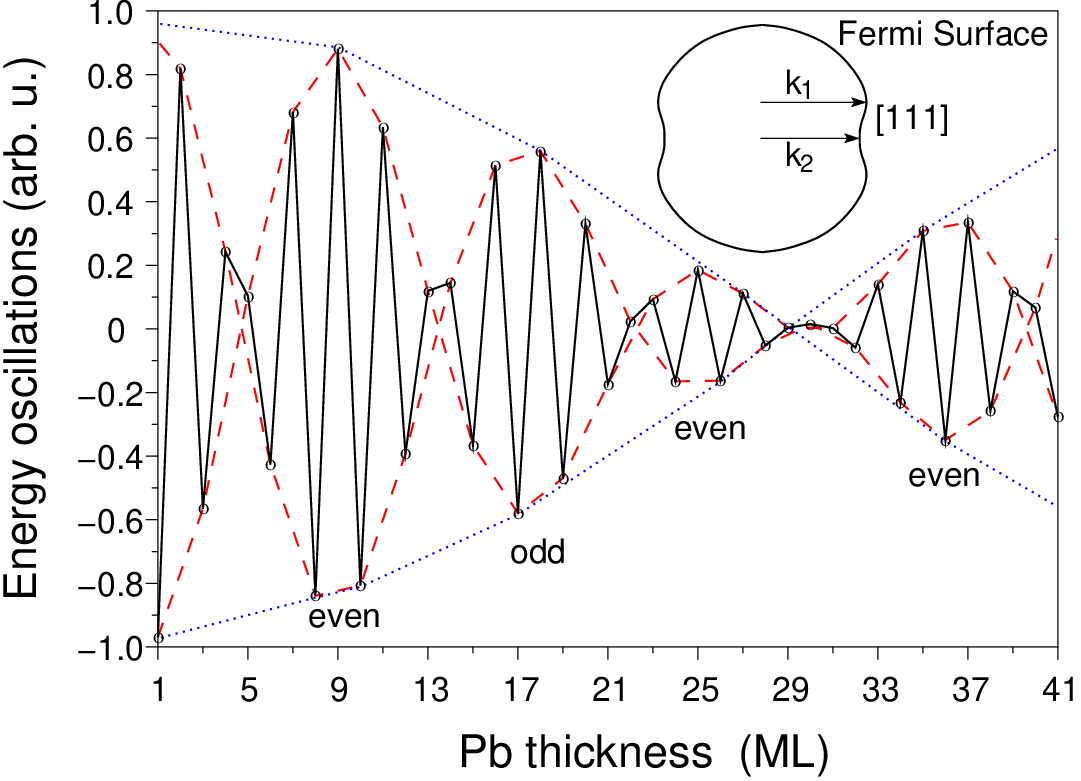}
\caption{{\it  (Color on line)}  Second derivative  of the  energy per
surface  atom times the  square of  the Pb  slab thickness  versus the
number  of  ML's  (continuous  line).  The dotted  blue  line  is  the
analytical  curve obtained  with two  values of  the  Fermi wavelength
$\lambda_{1}=7.47 a_0$ and  $\lambda_{2}=7.54 a_0$ with weights $A_1=0.72$
and $A_2=1.28$ respectively. The dashed line is its envelope function.}
\label{fig:beating_ab_initio}
\end{figure}

For   small  thickness  (   $N<30$)  the   structure  is   similar  to
Fig. \ref{fig:jell-ana},  so there  is no contradiction  with previous
simple  models.  However,  our central  predictions are  obtained when
looking  at  the periods  and  decay  of  the oscillations  at  larger
thickness.       The      amplitude      of     the      oscillations,
$-\frac{d^2\sigma}{dN^2}D^2$, in the  $N=1-40$ range decays, while it remains constant for the jellium data, i.e., the ab-initio results have
stronger damping  than the  jellium results.  Nevertheless,  the  amplitude for sizes larger than 40 ML remains constant. To  our knowledge,
experiments have not explored distances beyond this thickness, but our
results indicate that oscillations would survive for larger sizes than
explored currently up to now.

In addition,  a progressive shortening  of the beat periods  for large
thicknesses ($N>30$) is observed. There is a non-negligible period change
between N=33 and N=39 with a value of 6.35 ML.

These trends are reproduced analytically with a simple description as
the interference of two sinusoidal functions
\begin{equation}
 A_1\sin(2k_1(d    N+\delta_0))    +A_2\sin(2k_2(d
N+\delta_0))
\label{eqn:1}
\end{equation}
where $k_1$  and $k_2$ are the  two nesting Fermi wave  vectors in the
[111] direction (see sketch  in Fig. \ref{fig:beating_ana}), $A_1$ and
$A_2$ are  their corresponding weights, $\delta_0$ is  a surface shift
that accounts for the wavefunction spill out at the slab walls, $N$ is
the number of  atomic layers and $d$ is the  interlayer spacing in the
(111) direction.  The  ab-initio calculation is  in excellent agreement
with      the      analytical      model     (dotted      line      in
Fig. \ref{fig:beating_ab_initio}).  In fact, the key point is that our
ab-initio calculations  can not  be fitted using  only one  Fermi wave
vector. Note  that one Fermi wave  vector is enough to  fit the jellium
results of  Fig. \ref{fig:jell-ana}(b).  In summary,  an extra nesting
of the Fermi vectors is clearly involved in the mechanism to stabilise
the Pb slabs.

Next,  we   are  comparing  the   sum  of  sinusoidal   functions,  Eq.
\ref{eqn:1}, and the  experiments for supported Pb layers.  We use Eq.
\ref{eqn:1}   to   fit   the   experimental   results   of   Pb/Si(111)
\cite{czoschke}.    The   energy  oscillations   are   given  in   Fig.
\ref{fig:beating_ana}.   The weights  are $A_1=A_2=0.5$.   It  must be
noticed  that the  position of  the second  quantum beat  is inversely
proportional   to  the   difference   of  the   wavelengths,  so   its
determination is  very sensitive to  the input parameters.   The Fermi
wave length  of the used vectors  are very close to  the free electron
$\lambda_F$s  and  to  the  values  found in  the  literature  for  Pb
($\lambda_1=7.50$ $au$ and  $\lambda_2=7.59$ $au$ with an experimental
error  around 0.1  $au$ \cite{otero}).   The energy  oscillations have
beats  every  $\approx  8$ ML's,  as  it  is  known,  but with  a  new
modulation and the second quantum beat  is at 29 ML's.  The absence of
a clear  second quantum  beat in Fig.   \ref{fig:beating_ab_initio} is
because  $A_1\neq A_2$.   Although the  largest thickness  measured is
around  the  second  quantum  beat position,  our  fitting  reproduces
completely  the measured  features, both  oscillations  and amplitudes.
Again, the  agreement using only one  Fermi wave vector  is worse.  An
additional hint of the double  quantum beat comes out by the even-even
successive regions intercalated in the even-odd alternation pattern of
Fig.   \ref{fig:beating_ana}.  This  even-even  characteristic appears
about  $N=29$.  It  seems to  go  unnoticed in  the already  published
experimental  paper  \cite{czoschke}, where  they  have measured  with
X-ray diffraction  up to $N\approx  35$ MLs, and therefore  with lower
resolution  this even-even  pattern  emerges just  in the  borderline.
Further experimental  work is encouraged in  order to look  for such a
second quantum beat pattern.

\begin{figure}
\includegraphics[width=\columnwidth]{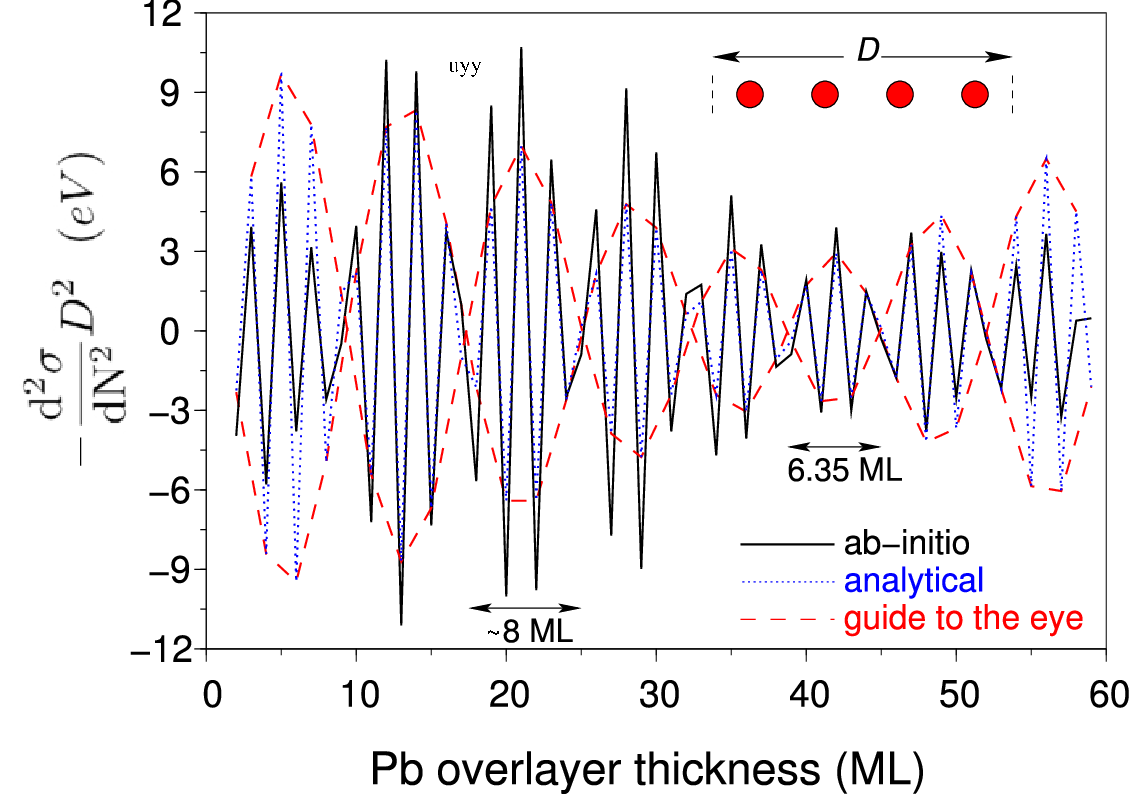}
\caption{  Modulated oscillatory  pattern  of the  energy obtained  by
superposition  of  two oscillations  following  Eq. (\ref{eqn:1})  and
using  the   Fermi  wavelength  values   $\lambda_1=7.40$  $a.u.$  and
$\lambda_2=7.48$ $a.u.$, that are obtained by fitting the experimental
curve  \cite{czoschke,two}.  The dashed  line is  the envelope  of the
first beat pattern  and the dotted line is the  envelope of the second
modulation, the so-called second beat  pattern.  The inset is a sketch
of the cross-section of the  Fermi surface of Pb, showing both nesting
vectors in (111) direction.}
\label{fig:beating_ana}
\end{figure}

%Summing up, we extract the  values of the nesting wavelengths underlying
%the  oscillations  obtained  from  our  ab-initio  calculations  by  a
%spectral analysis.  The Fourier  transforms of the energy oscillations
%of  previous figures  are shown  in Fig.  \ref{fig:fourier}.  Although
%Fig.  \ref{fig:fourier}, concerning the analytical results, supplies a
%rather  accurate estimate  of the  number  of spectral  peaks and  the
%frequencies  at  which  they  occur, for  ab-initio  calculations  the
%information regarding intensities and lines shapes is broadened due to
%the  actual  window  of  layer   sizes.  We  note  that  both  beat
%frequencies       for        the       analytical       model       of
%Fig. \ref{fig:beating_ab_initio} have merged into a broad peak as seen
%in the  inset of Fig.   \ref{fig:fourier}. This broad  pattern appears
%also in the atomistic calculations in  a similar way and with the same
%values on the right side of this plot.

%\begin{figure}
%\includegraphics[width=\columnwidth]{Fig4.eps}
%\caption{Frequency  for   the  previous  oscillations   after  Fourier
%transformations. First-principle calculations  are given with the full
%lines;  analytical expressions  for the  one and  two  wavelengths are
%shown  in the  dot and  dashed lines  respectively. In  the  plot, the
%window of the analytical transforms  is three time the one of ab-initio
%calculations. }
%\label{fig:fourier}
%\end{figure}

The  decay  of  the amplitude  of  the  energy  smear out  the  energy
oscillations  at  large  thickness,   so  the  size  selection  should
disappear in experiments.  Nevertheless, if this is accompanied by the
second quantum  modulation we predict  that they must emerge  again at
larger sizes. The observation of this effect in the stability is still
an  open question  because  the  experiments did  not  go beyond  that
size. For example,  the same effect of the  non-spherical shape of the
Fermi  surface has  been also  pointed out  to influence  the magnetic
interlayer coupling in multilayers,  being important upto distances as
long  as 100  ML's  \cite{bruno,holmstrom}.  This  indicates that  the
detection  of  high  stability   with  larger  sizes  is  possible  in
experiments.
% Anyhow,  the shortening of the period  obtained from our
%ab-initio calculations is located after the maximum number of ML's (35
%ML)  actually  used in  the  experiments  \cite{upton}.  

%Second,  when
%studying realistic layers on surfaces we have to take into account the
%spillage between  the Pb layers and  the substrate \cite{upton,otero}.
%At the  low-coordinated interface the  s-p charge spills out  from the
%ion core regions lowering the  kinetic energy. The charge spill-out in
%turn  increases  the local  work  function,  resulting  in the  charge
%transfer  between  the  layers  and  the substrate  in  the  direction
%determined by the relative positions  of the work function. Thus, this
%determines the relative shift  of the different beats instead of
%the given patterns for free standing  Pb.  We can see this shift by
%comparing  our first beat  node with  calculations on  a substrate,
%such  as Pb/Ge(111)  \cite{materzanini}, or  just relaxing  the layers
%\cite{ozer}, both  using few Pb  monolayers.  Then, our  above results
%regarding the  beats of the analytical expressions  can be applied
%to the case of depositing layers on surfaces only qualitatively, although all
%%the  free parameters  are  fitted or  calculated properly.   Moreover,
%first-principles calculations for Pb  layers on silicon surfaces as in
%some experimental samples are under progress.\\

In  conclusion, both  model and  density functional  calculations show
that there  is a reentrance of  QSE's and ensuing new  features in the
stability  patterns  at  sizes  substantially larger  than  the  sizes
studied  up to  now (or  in the  borderline). This  suggests  that the
nesting  with a second  wave vector  of the  Fermi surface  provides a
mechanism to understand the stability at the large sizes in the growth
of Pb layers. This second quantum beat claims for a new interpretation
of experiments, as well as for new measurements at higher Pb coverages
to shed light about the existence of the second quantum modulation and
its role in the self-selection and self-assembly processes. \\

{\bf Acknowledgements:} 

 This  work was  supported by  the  ETORTEK( NANOMAT)  program of  the
Basque government, Spanish Ministerio de Ciencia y Tecnolog\'ia (MCyT)
of Spain( Grant No. Fis 2004-06490-CO3-00) and the
European Network  of Excellence NANOQUANTA  (NM4-CT-2004-500198).  The
SGI/IZO-SGIker  UPV/EHU (supported  by  the National  Program for  the
Promotion of  Human Resources within  the National Plan  of Scientific
Research, Development and Innovation  - Fondo Social Europeo, MCyT and
Basque  Government)  is  gratefully  acknowledged  for  allocation  of
computational resources.


\begin{references}

\bibitem{genzken}
O. Genzken and M. Brack, Phys. Rev. Lett. 67, 3286 (1991).

\bibitem{yanson1999} A.  I.  Yanson, I.K.
Yanson, and J.   van Ruitenbeek,  Nature 400, 144  (1999).

\bibitem{yanson2000}
A. I.  Yanson, I.K. Yanson, and
J.  van Ruitenbeek, Phys. Rev. Lett. 84, 5832 (2000).

\bibitem{yanson2001}
A. I.  Yanson, I.K. Yanson, and
J.  van Ruitenbeek, Phys. Rev. Lett. 87, 216805 (2001).


\bibitem{luh}
D.-A. Luh, T. Miller, J. J. Paggel, M. Y. Chou, and T.-C. Chiang,
Science {\bf 292}, 1131 (2001).

\bibitem{floreano}
L. Floreano, D. Cvetko, F. Bruno, G. Bavdek, A. Cossaro, R. Gotter, A. Verdini and A. Morgante,
Progress in Surf. Sci. {\bf 72}, 135 (2003).

\bibitem{aballe}
L. Aballe, C. Rogero and K.Horn,
Surf. Sci. {\bf 518}, 141 (2002).

\bibitem{saito}
M. Saito, T. Ohno, and T. Miyazaki,
Applied Surf. Sci. {\bf 237}, 80 (2004).
\bibitem{upton}
M.H. Upton, C.M. Wei, M.Y. Chan, T.Miller, and T.C.Chiang, Phys. Rev.
Lett. {\bf 93},026802 (2004).

\bibitem{czoschke}
P. Czoschke, H. Hong, L. Basile, and T.-C. Chiang, Phys. Rev. Lett.
{\bf 93}, 036103 (2004).

\bibitem{otero}
R. Otero, A. L. V\'azquez de Parga, and R. Miranda, Phys. Rev. B {\bf 66},
115401 (2002).

\bibitem{hupalo}
M. Hupalo and M.C. Trigides,
Phys. Rev. B  {\bf 65}, 115406 (2002).

\bibitem{schulte}
F.K. Schulte,
Surf. Sci. {\bf 55}, 427 (1976).

\bibitem{zhang}
Z. Zhang, Q. Niu, and C.-K. Shih, Phys. Rev. Lett. {\bf 80}, 5381 (1998).

\bibitem{edu1}
E. Ogando, N. Zabala, E.V. Chulkov and M.J. Puska,
Phys. Rev. B  {\bf 69}, 153410 (2004).

\bibitem{edu2}
E. Ogando, N. Zabala, E.V. Chulkov and M.J. Puska,
Phys. Rev. B  {\bf 71}, 205401 (2005).

\bibitem{parkin}
S.  S.  P.  Parkin, R.  Bhadra,  and K.  P.  Roche,
Phys.  Rev.  Lett.   66, 2152 (1991).

\bibitem{johnson}
M.  T.  Johnson,  S. T.  Purcell,
N.  W.  E. McGee, R. Coehoorn,  J.  Van de Stegge, and W. Hoving, Phys.
Rev.  Lett.  68,  2688 (1992).

\bibitem{bruno1991}
P.  Bruno and  C.  Chappert, Phys.  Rev.
Lett.  67, 1602 (1991).

\bibitem{edwards}
D.  M.  Edwards,J.  Mathon, R.  B.  Muniz, and
M. S. Phan, Phys. Rev. Lett.  67, 493 (1991).

\bibitem{alonso}
J.A. Alonso and M. J. Lopez, Journal of Cluster Science, {\bf 14}, 
31 (2003).

\bibitem{anderson}
J.R. Anderson and A.V. Gold, Phys. Rev. {\bf 139}, A1459 (1965).


\bibitem{vasp}
G. Kresse and J. Furthm\"uller, Comput. Mater. Sci. {\bf 6}, 15
(1996); Phys. Rev. B  {\bf 54}, 11 169 (1996).

\bibitem{perdew}
J. Perdew, K. Burke, and M. Ernzerhof, Phys. Rev. Lett. 77, 3865 (1996).

\bibitem{kresse} 
G. Kresse, and D. Joubert, Phys. Rev. B {\bf 59}, 1758 (1999).

\bibitem{details} Convergence of the energies versus plane-wave cutoff
(237  eV)  , k-mesh  (22x22),  and vacuum  (10  ML)  were chosen  when
requiring  high  accuracy. These  values  are  in  agreement with  the
reference D. Yu and M. Sheffler, Phys. Rev. B {\bf 70}, 155417 (2004).



\bibitem{wei} 
C.M. Wei, M.Y. Chou, Phys. Rev. B {\bf 66}, 233408 (2002).

\bibitem{materzanini}
G. Materzanini, P.  Saalfrank, and
P.J.D.  Lindan, Phys. Rev. B 63, 235405 (2001).

\bibitem{bruno}
P. Bruno, J. Phys.: Condens. Matter. {\bf 11}, 9403 (1999).


\bibitem{holmstrom}
E. Holmstr\"om, A. Bergman, L. Nordstr\"om, I.A. Abrikosov, S.B. Dugdale and B.L. Gy\"orffy, Phys. Rev. B {\bf 70}, 64408 (2004). 


\end{references}
\end{document}